\documentclass{sig-alternate-10pt} 
\usepackage{color}
\usepackage{multirow}
\usepackage{url}
\usepackage{algorithm}
\usepackage[noend]{algpseudocode}
\usepackage{enumitem} 
\usepackage{subfigure}

\newcommand{\myitem}[1]{\vspace*{0.07in}\noindent\textbf{#1}}
\newcommand{\afteritem}{\vspace*{0.07in}}
\newcommand{\xin}[1]{\textcolor{red}{(XIN: #1)}}

\definecolor{Ora}{cmyk}{0, 0.6, 0.8, 0}
\newcommand{\lv}[1]{\textcolor{Ora}{(LV: #1)}}
\newcommand{\cut}[1]{}

\title{SoftCell: Taking Control of Cellular Core Networks}

\numberofauthors{1}
\author{
Xin Jin$^\dagger$, Li Erran Li$^\star$, Laurent Vanbever$^\dagger$, and Jennifer Rexford$^\dagger$\\
Princeton University$^\dagger$, Bell Labs$^\star$}

\begin{document}
\maketitle
\thispagestyle{empty}


\begin{abstract}
Existing cellular networks suffer from inflexible and expensive equipment,
and complex control-plane protocols.  To address these challenges,
we present SoftCell, a scalable architecture for supporting
fine-grained policies for mobile devices in cellular core networks.
The SoftCell controller realizes high-level service polices by
directing traffic over paths that traverse a sequence of middleboxes,
optimized to the network conditions and user locations.  To ensure
scalability, the core switches forward traffic on hierarchical
addresses (grouped by base station) and policy tags (identifying paths
through middleboxes).  This minimizes data-plane state in the core
switches, and pushes all fine-grained state to software switches at
the base stations.  These access switches apply fine-grained rules,
specified by the controller, to map all traffic to the appropriate
addresses and tags.  SoftCell guarantees that packets in the same
connection traverse the same sequence of middleboxes in both
directions, even in the presence of mobility.  Our characterization of
real LTE workloads, micro-benchmarks on our prototype controller, and
large-scale simulations demonstrate that SoftCell improves the
flexibility of cellular core networks, while enabling the use of
inexpensive commodity switches and middleboxes.
\end{abstract}


\section{Introduction}
\label{sec:intro}
\begin{figure}[t]
\centering
\vspace*{-0.35in}
\includegraphics[width=3.3in]{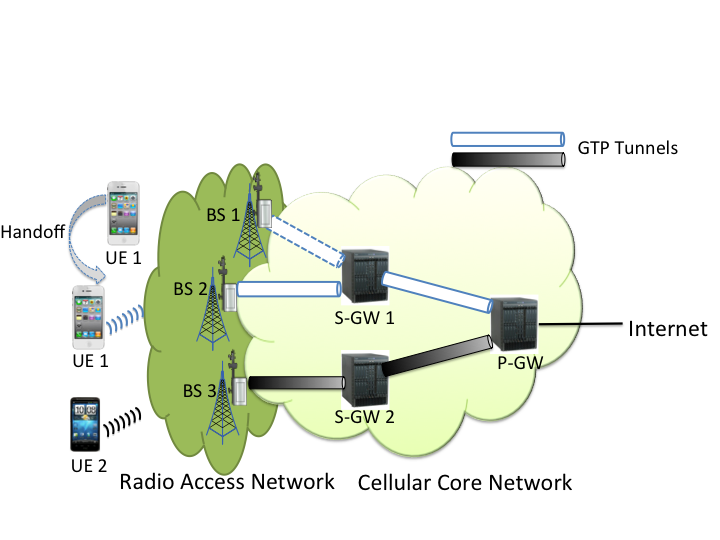}
\vspace*{-0.45in}
\caption{LTE network architecture}
\vspace*{-0.15in}
\label{fig:lte-dataplane}
\end{figure}

The rapid proliferation of cellular devices (e.g., smart phones,
tablets, and smart meters) is pushing existing cellular networks to
their limits.  New technologies like Long Term Evolution (LTE) are
helping increase the capacity of radio access networks, placing even
greater demands on cellular core networks to support many diverse
devices and applications.  Cellular core networks carry traffic
between base stations and the Internet on behalf of user equipment
(UE), as shown in Figure~\ref{fig:lte-dataplane}.  The network relies
on specialized equipment such as serving gateways (S-GWs) that provide
seamless mobility when UEs move between base stations, and packet
gateways (P-GWs) that perform a wide variety of functions like traffic
monitoring and billing, access control, and parental controls.  The
base stations, serving gateways, and packet gateways communicate over
GTP tunnels traversing a network of switches and routers.

Cellular core networks are remarkably complex and
inflexible~\cite{cell-directions,elby-sdn}, an unfortunate legacy of
their circuit-switched origins.  Centralizing critical data-plane
functionality at the boundary with the Internet forces all traffic to
flow through the packet gateway---including device-to-device traffic
and local Content Distribution Network (CDN) services within the same
cellular network.
With so much functionality in one box, it is not surprising that
packet gateways are complex and expensive, and force carriers to buy
functionality they do not need.  Carriers cannot ``mix and match''
capabilities from different vendors (e.g., a firewall from one vendor,
and a transcoder from another), or ``scale up'' the resources devoted
to a specific function~\cite{elby-sdn,nfv}.  Since the packet gateways
are hard to change, carriers are forced to replace them to deploy new
functionality, even when the existing equipment suffices for most
purposes.

To make matters worse, growing link speeds and more diverse network
policies will put even greater strain on packet gateways in the
future.  Cellular networks can apply customized policies based on a
wide variety of \emph{subscriber attributes} (e.g., the cell-phone
model, the operating-system version, the billing plan, options for
parental controls, whether the total traffic exceeds a usage cap, and
whether a user is roaming), as well as the \emph{application} (e.g.,
transcoding for video traffic, caching for Web traffic, and exemption
from usage caps for applications that pay the carrier on the user's
behalf)~\cite{elby-sdn}.  For example, the carrier may direct traffic
for older cell phones through an echo-cancellation gateway, video
traffic through a transcoder during times of congestion, and all
traffic through a firewall, while applying different monitoring
policies depending on the billing plan, usage cap, roaming status, and
the application.

Rather than perform all these functions at the Internet boundary, we
argue that cellular providers should adopt a network design more akin
to modern data centers.  The network should consist of a fabric of
simple core switches, with most functionality moved to
low-bandwidth access switches (at the base stations) and a distributed
set of middleboxes that the carrier can expand as needed to
meet the demands.  These middleboxes could be dedicated appliances,
virtual machines running on commodity servers~\cite{nfv}, or simply
packet-processing rules installed in the
switches~\cite{ethane,load-wild}.  A logically-centralized controller
can direct traffic through the appropriate middleboxes, via efficient
network paths, to realize a high-level service policy (e.g., directing
a UE's video traffic through a transcoder and a firewall).

Cellular networks raise unique scalability challenges, compared to
data-center and enterprise networks.  Fine-grained policies can easily
lead to an explosion in the data-plane state needed to direct traffic
through the right middleboxes.  This is especially true for the large
volume of ``north-south'' traffic arriving from the Internet.  In
addition, stateful middleboxes require that all traffic in the same
connection traverses the same middleboxes, even when a UE moves from
one base station to another.  The switches need to forward packets
differently based on multiple factors (e.g., the UE and the
application), which typically requires expensive TCAM (Ternary Content
Addressable Memory) for packet classification.  However, the merchant
silicon chipsets used in commodity switches have just a few thousand
to tens of thousands of TCAM entries.  (See Table 2 in~\cite{past}.)
Supporting much larger packet classifiers would significantly increase
the cost of the core switches.

To address these challenges, we present SoftCell, a scalable
architecture for supporting fine-grained policies for mobile devices
in cellular core networks.  The SoftCell controller realizes
high-level service polices by directing traffic through a sequence of
middleboxes, optimized to the network conditions and UE locations.  To
ensure data-plane scalability, the core switches forward traffic on
\emph{hierarchical addresses} (grouped by base station) and \emph{policy tags}
(identifying middlebox paths).  SoftCell pushes fine-grained packet
classification to the access switches, which can be implemented easily
in software.  These access switches apply fine-grained rules,
specified by the controller, to map UE traffic to the policy tags and
hierarchical addresses.  To ensure control-plane scalability, a local
agent at the base station caches the service policy for each attached
UE, to install rules in the access switch without involving the
controller.

The SoftCell controller guarantees that packets in the same connection
traverse the same sequence of middleboxes (\emph{policy consistency}),
and that bidirectional traffic traverses the same middleboxes in both
directions (\emph{policy symmetry}), even in the presence of mobility.
SoftCell has an \emph{asymmetric} edge architecture that does
\emph{not} require sophisticated packet classification of return
traffic arriving at the gateway switches.  SoftCell either
\emph{embeds} the policy tags in the UE IP address and port number
(essentially ``piggybacking'' the information in the packets sent to
the Internet), or \emph{caches} them at the gateway (in a simple
Network Address Translation table).  This ensures return traffic flows
through the right middleboxes, without requiring any support from the
rest of the Internet.  SoftCell also does not require any changes to
UEs or the radio access network hardware, and can run on commodity
switches and middleboxes.

In designing, prototyping, and evaluating SoftCell, we make the
following contributions:

\myitem{Fine-grained service polices:} SoftCell supports fine-grained traffic
steering based on applications and subscriber attributes, as well as
flexible traffic engineering in selecting the network and middlebox paths.

\myitem{Asymmetric edge design:} SoftCell places most functionality
at the many, low-bandwidth access switches, allowing the core network
to use commodity hardware for the Internet gateway and other core
switches.

\myitem{Scalable data plane:} SoftCell minimizes data-plane state in
the core switches through multi-dimensional aggregation by policy
tags, base station IDs, and UE IDs, and an algorithm for selecting policy tags.

\myitem{Scalable control plane:} To ensure control-plane scalability,
access switches run local agents that cache service policies for the
attached UEs, and the controller isolates the access switches from
core topology changes.

\myitem{Policy consistency and symmetry:} SoftCell ensures that all
traffic in the same TCP or UDP connection traverses the same sequence
of middleboxes in both directions, even in the presence of mobility.

\myitem{Realistic performance evaluation:} We evaluate the scalability
our architecture based on traces from a large LTE deployment,
micro-benchmarks on a prototype controller, and large-scale simulation
experiments.
\afteritem

We believe SoftCell significantly improves the flexibility of cellular
core networks, while enabling the use of inexpensive commodity
switches and middleboxes.


\section{SoftCell Architecture}
\label{sec:arch}
A SoftCell network consists of commodity middleboxes and switches
managed by a controller.  The controller supports flexible, high-level
service policies by computing and installing rules in the switches to
direct traffic through the right middleboxes and network paths.  To
support flexible policies without compromising scalability, SoftCell
capitalizes on the unique properties of cellular core
networks---particularly the fact that most traffic begins at the
base-station edge, where the small number of flows and the small
uplink bandwidth enable the use of flexible software switches.

\subsection{SoftCell Core Network Components}
\begin{figure}
\centering
\includegraphics[width=3.3in]{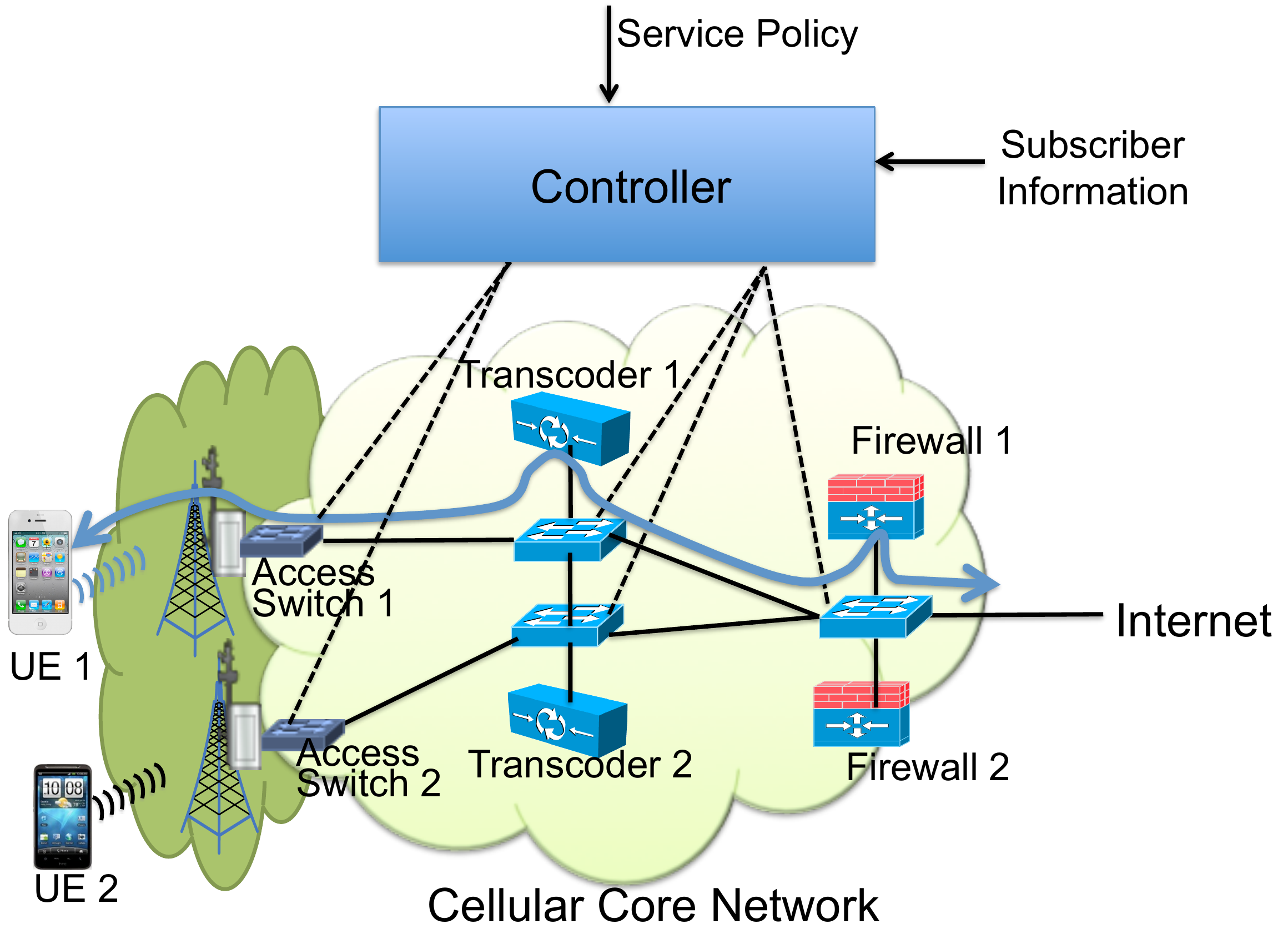}
\vspace*{-0.45in}
\caption{SoftCell network architecture}
\vspace*{-0.15in}
\label{fig:high-level}
\end{figure}

The cellular core network connects to unmodified 
UEs (via base stations) and the Internet (via gateway switches), as shown
in Figure~\ref{fig:high-level}.  SoftCell does \emph{not} require the
specialized network elements (e.g., serving and packet gateways) or
point-to-point tunneling (e.g., user-level GTP tunnels) used in
today's LTE networks, as shown earlier in Figure~\ref{fig:lte-dataplane}.

\myitem{Middleboxes:} SoftCell supports commodity middleboxes
implemented as dedicated appliances, virtual machines, or
packet-processing rules on switches.  Each middlebox function (e.g.,
transcoder, web cache, or firewall) may be available at multiple
locations.  Many middleboxes require all packets in both directions
of a connection to traverse the same instance of the middlebox.

\myitem{Access switches:} Each base station has an \emph{access}
switch that performs fine-grained packet classification on traffic
from UEs.  Access switches can be software switches (such as Open
vSwitch~\cite{openvswitch}) that run on commodity server hardware.
The server can also run a local agent that caches service policies for
attached UEs, to minimize interaction with the central controller.

\myitem{Core switches:} The rest of the cellular core consists of
\emph{core} switches, including a few \emph{gateway} switches
connected to the Internet.  These core switches perform
multi-dimensional packet classification at high speed, but only for a
few thousands or tens of thousands of rules.  We assume that the
packet-processing hardware can perform arbitrary wildcard matching on
the IP addresses and TCP/UDP port numbers (as in today's merchant
silicon), or can cache flat rules after processing wildcard rules
locally in software (as in DevoFlow~\cite{devoflow}).
Our gateway switches are much cheaper than P-GWs. They can
be flexibly placed at many locations with access to the Internet.   
SoftCell enables a \emph{``flatter'}' core network
architecture with more efficient routing than current LTE does. 

\myitem{Controller:} The controller computes and installs switch-level
rules that realize a high-level service policy, specified based on
subscriber attributes and applications, by installing paths that
direct traffic through middleboxes. The controller knows the
attributes (e.g., billing plan, phone model, and usage cap) of each
UE, allowing the controller to identify the appropriate clauses in
the service policy for handling the UE's traffic.  
\afteritem

The radio access networks consist of base stations that connect to
unmodified UEs using existing protocols for mobility management,
session management, and authentication.  Just as today, a UE retains a
single IP address as it moves between base stations in the same
cellular core network; any changes our cellular core network makes to
the IP addresses of packets are not visible to the UEs.  We do not
change the radio hardware at the base station, or common functions
such as scheduling, radio resource management, and paging.  SoftCell
only changes how the base stations communicate with the core network,
by having the base stations coordinate with the controller to enforce
service policies.  Similarly, SoftCell does not require changes to
commodity middleboxes, or any support from the rest of the Internet.

\subsection{Flexible, High-Level Service Policies}
The SoftCell controller directs traffic over network and middlebox
paths, based on the service policy. We believe that carriers should
specify service policies at a high level of abstraction, based on
subscriber attributes and applications, and rely on the controller to
handle low-level details like ephemeral network identifiers, the
locations of middleboxes and switches, and application
identification. A service policy has multiple clauses that each
specify which traffic (specified by a predicate) should be handled in
what way (specified by an action):

\myitem{Predicates:} A predicate is a boolean expression on subscriber
attributes, application type, and cell properties. Subscriber
attributes consist of device type, billing plan, device capabilities,
provider, etc.  Application types include web browsing, real-time
streaming video, VoIP, etc.  Cell attributes include the air interface
congestion level, capacity, etc.

\myitem{Service action:} An action consists of a set of middleboxes,
along with quality-of-service (QoS) and access-control specifications.
Specifying the set of middleboxes as a partial order allows the
carrier to impose constraints (e.g., firewall before transcoder).  The
action does not indicate a specific instance of each middlebox,
allowing the controller to select middlebox instances and network
paths that minimize latency and load.

\myitem{Priority:} The priority is used to disambiguate overlapping
predicates.  The network handles traffic using the highest-priority
clause with a matching predicate.
\afteritem

\begin{table}
\centering
{\small  
\begin{tabular}{|c|l|l|} \hline
\multicolumn{1}{|c}{\bf Pri} & 
  \multicolumn{1}{|c}{\bf Predicates} &
  \multicolumn{1}{|c|}{\bf Service Action} \\ \hline
1 & provider = B & Firewall \\  \hline
2 & provider != A & Drop  \\  \hline
3 & app = video~$\wedge$~plan = Silver & [Firewall, Transcoder] \\
   & \hspace*{0.1in} $\wedge$ congestion > 7 & \\ \hline
4 & app = VoIP & QoS = expedited-forward\\
   & & \hspace*{0.1in} $\wedge$ Firewall \\
\hline
5 & $*$ & Firewall \\ 
\hline
\end{tabular}
}
\vspace*{-0.1in}
\caption{Example service policy for carrier $A$}
\vspace*{-0.15in}
\label{tab:expol}
\end{table}

Table~\ref{tab:expol} shows an example service policy that carrier $A$
applies to traffic arriving at UEs, where outbound traffic follows the
reverse sequence of middleboxes.  Carrier $A$ has a roaming agreement
with carrier $B$, so the first clause directs traffic from $B$'s
subscribers through a firewall.
The second clause disallows traffic from subscribers from all other
carriers.  The remaining clauses specify the handling of $A$'s own
subscribers, with all traffic going through a firewall.  The third
clause indicates that the video traffic to subscribers on the
``silver'' billing plan must go through a transcoder (after the
firewall) when cell congestion at the base station exceeds a target
level.
The fourth clause specifies that VoIP traffic should be assigned to
the ``expedited forwarding'' service class to protect this application
from a heavy load of best-effort traffic.  The fifth clause requires
that all other traffic goes through a firewall.  In this paper, we
focus on middlebox service policies, since they require more
sophisticated \emph{traffic steering} rather than simple local
processing to drop packets or mark the type-of-service bits.

\subsection{Scalability Design Principles}
The main challenge in SoftCell is to support flexible policies without
compromising scalability.  To motivate our main design decisions, we
briefly discuss the main factors affecting the scalability of cellular
core networks, and perform back-of-the-envelope calculations based on
publicly-available statistics. We consider the design of a typical
cellular core network serving a large metropolitan area with 1000
base stations~\cite{celltopology}.

\myitem{Microflow rules in software access switches:}
%
A modern base station can serve around 1000 UEs~\cite{eNodeBALU}. Not
surprisingly, most UEs have just a handful of active TCP or UDP
connections at a time~\cite{rahmati,zhang12}.  A base station with
1000 UEs, each with (say) 10 active TCP/UDP connections, would have 10K
simultaneously active flows.  The backhaul link from the base station
to the rest of the core network has a capacity of anywhere from 20 Mbps
to 1 Gbps~\cite{celltopology,Ericsson1G}.  A software switch like Open
vSwitch~\cite{openvswitch} can easily store 100K microflows in a hash table, and
perform packet forwarding at several gigabits per second~\cite{ovs-perf},
comfortably within these requirements.  Open vSwitch can install around 1K flow
entries every 100 msec~\cite{oflops}, able to support a new
flow from each UE every tenth of a second.  These results suggest that
a software switch can easily keep up with the number of flows, flow
set up rates, and aggregate bandwidth at each base station.

\myitem{Exploit the dominance of UE-initiated traffic:}
Most traffic in cellular networks is ``north south", as opposed to
data-center networks where most traffic is ``east west".  What's
more, clients are usually (almost always) UEs in these ``north south"
traffic, which means traffic is first initiated from UEs.  Actually,
many cellular operators deploy NATs and stateful firewalls to forbid
connections initiated from the Internet \cite{wang2011untold}, as a
way to protect their networks.  This factors into our solution to
place most key functionality at access switches.

\myitem{Avoid fine-grained packet classifiers at the gateway switches:}
The gateway switches need to handle the traffic for 1000 base
stations, each with (say) 10K active flows.  This results in roughly
10 million active flows---too large for fine-grained packet classification using commodity switch hardware. To
leverage merchant silicon with thousands to tens of thousands of TCAM
entries, the data-plane state in the core switches should not be more
than (say) an order of magnitude higher than the number of base
stations.  As such, the gateway switches should \emph{not} perform
fine-grain packet classification to identify the base station or
service action associated with the incoming packets.

\myitem{Exploit locality to reduce data-plane state in the core
  switches:}
%
Fortunately, cellular core networks have natural geographic locality,
with the access switches aggregating through metro networks to mobile
switching offices, through the core to gateway switches.  Since a \emph{cluster}
of around $10$ base stations connect (in a ring, tree, or mesh
topology) to the core \cite{bscluster}, aggregating by base station clusters can
reduce data-plane state by an order of magnitude.  In addition,
traffic for these base stations would often traverse the same nearby
middlebox instances (to minimize latency and network load), offering
further opportunities to reduce the state required to support service
policies.

\myitem{Avoid fine-grained events at the controller:} While the access
switches can maintain per-flow state, the \emph{controller} cannot manage the
network at the flow level.  With an arrival rate of (say) $1$-$10$K
flows/second from each of 1000 base stations, a controller that
processes microflows would need to handle $1$M-$10$M events per
second, roughly doable with today's SDN controller
platforms~\cite{controller-perf}, but only at the expense of flow
set-up latency and high overhead.  Instead, we believe the SoftCell
controller should only handle coarse-grained events, such as UE
arrivals at base stations or traffic requiring a new service action.

\myitem{Avoid updating access switches after topology changes:}
%
In addition to satisfying the service policy, the controller must
response in real time to network events such as link failures or
congestion by computing and installing new rules in the switches.  If
the controller also needed to update all 1000 base stations, routing
convergence time would suffer.  Instead, routing changes should be
isolated to the core switches, without updating the access switches.
\afteritem

The first four principles ensure that SoftCell has a scalable data
plane, as discussed in Section~\ref{sec:data}.  Then,
Section~\ref{sec:control} applies the last two principles to ensure
the control plane scales.


\section{Scalable Data Plane}
\label{sec:data}
%



\begin{figure*}[t]
\centering
\includegraphics[width=6.in]{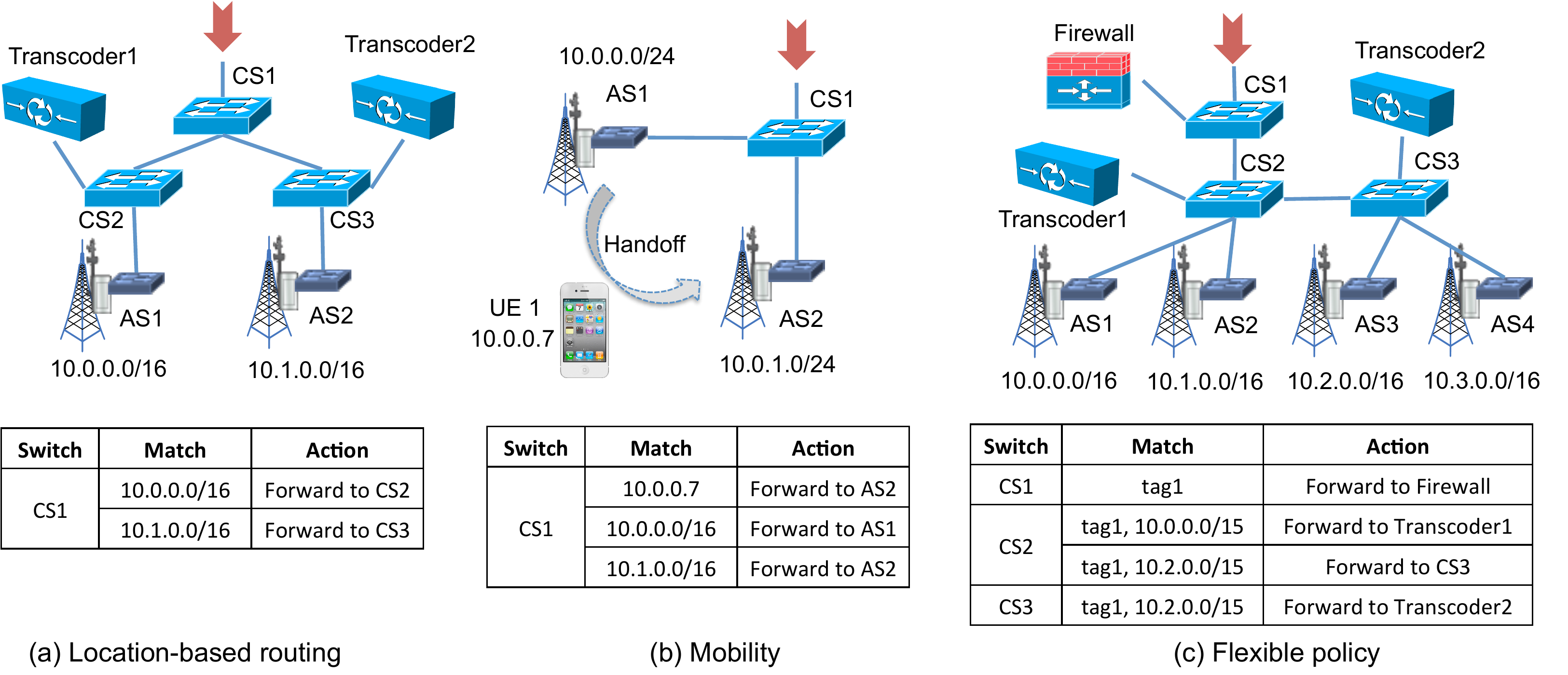}
\vspace*{-0.1in}
\caption{Examples of multidimensional aggregation rules for traffic arriving from the Internet}
\label{fig:ex}
\end{figure*}

To ensure the scalability of the data plane, the access switches apply
fine-grained rules that map packets to hierarchical addresses and
coarse-grained policy tags, with the help of the controller.  The core
switches direct traffic based on these large aggregates.  By selecting
base station address blocks and policy tags intelligently, the
controller can enable aggregation across nearby base stations and
related policy tags to further reduce the state.  To avoid classifying
packets arriving from the Internet, SoftCell either embeds the
forwarding information in the IP+TCP/UDP header (essentially ``piggybacking''
the state in outgoing packets) or caches policy tags at the gateway.
When a UE moves from one base station to another, the controller
installs temporary rules in the core switches to direct in-progress
flows to the new location while ensuring policy consistency.

The controller directs traffic over a \emph{policy path}, a sequence
of switches and middleboxes from one edge switch to another.  To
simplify the discussion, we initially assume that the controller
handles the first packet of each flow, similar in spirit to
Ethane~\cite{ethane}.  This clearly would compromise the scalability
of the controller---an issue we address in Section \ref{sec:control}.

\subsection{Core: Multi-Dimensional Aggregation}
Delivering different traffic over different sequences of middleboxes is
hard to achieve in a scalable way. Suppose we have 1000 base stations,
each with 1000 UEs, where each UE has 1000 service
policy clauses. Installing a path for each service policy clause would lead to 1 billion
paths. If implemented naively, this would generate a huge amount of
rules in the switches. The key idea to achieve scalability is to aggregate
traffic on multiple dimensions, i.e., policies, base stations, and UEs.

\myitem{Aggregation by policy (policy tag):} Service policies defined
on high-level attributes seem very compact. However, subscriber
attributes are not easily translated or aggregated with network
addresses. For example, since UEs with ``Silver Plan" can have a
variety of IP addresses, the third clause of the service policy in
Table~\ref{tab:expol} may require a rule for each flow in the worst
case. We could conceivably assign ``Silver Plan" UEs IP addresses
under the same subnet, allowing us to assign one rule that matches on
the IP prefix.  However, we cannot do this for every attribute, not to
mention that many service policies are defined on combinations of
attributes. To minimize the rules in core switches, we use a
\emph{policy tag} to aggregate flows on the same policy path. We
associate packets with a policy tag at the access switch, allowing
core switches to forward packets based on coarse-grained policy tags.

\myitem{Aggregation by location (hierarchical IP address):} In many
core switches, traffic destined to the same base station would
traverse the same output link, even if the packets go through
different middleboxes. By including location information in the UE
addresses, we can aggregate traffic by IP prefix.  Furthermore,
cellular core networks have a natural hierarchical structure.
Therefore, we assign each base station an IP prefix, called \emph{base station ID},
and IDs of nearby base stations can be further aggregated into larger blocks. We can
aggregate even more by combining policy tags and IP addresses. Suppose
two policy paths going to two base stations share a long path segment
before branching.  If assigned the same policy tag, a single rule
matching on the tag can forward packets along the shared segment
until the branching point, where traffic divides based on the base
station prefix.

\myitem{Aggregation by UE (UE ID):} Packets also need a UE identifier (\emph{UE ID})
that differs from other UEs at the same base station. For example,
some middleboxes (like intrusion detection systems) need a way to
identify groups of flows associated with the same UE which is impossible if
all flows for the same base station share the same address.  In
addition, having a UE ID in each packet enables optimizations for
handling mobility, by installing switch rules that forward in-progress
flows to the UE at its new location. Together, the base station prefix
and the UE ID form a hierarchical location-dependent address (LocIP)
for the UE.  Using hierarchical ``care of" addresses to handle
mobility is an old idea~\cite{HAWAII02,CellularIP99}. However, prior
work does not consider service policies, or the techniques described
in the next two subsections to ensure policy symmetry and consistency.
\afteritem

Our key idea is to \emph{selectively} match on the three dimensions to
maximize aggregation of data-plane state:

\textbf{Location-based routing:} In Figure~\ref{fig:ex}(a), core switch CS1
matches on the base-station prefix to forward traffic to CS2 and
CS3. CS2 and CS3 decide whether to direct traffic to a transcoder
based on the policy tag, but CS1 does not need to base its forwarding decision on the tag.

\textbf{UE mobility:} In Figure~\ref{fig:ex}(b), CS1 forwards traffic
to base stations based on the destination IP prefix. When UE1 moves from access switch AS1 to
AS2, we install a high-priority rule at CS1 to match on both the base
station prefix and the UE ID. This ensures that ongoing flows reach
UE1 at AS2 over a direct path.

\textbf{Flexible policy:} Figure~\ref{fig:ex}(c) illustrates how to
implement the third clause in Table~\ref{tab:expol}, using the tag
``tag1.''  CS1 forward ``tag1'' packets to the
Firewall\footnote{Traffic \emph{from} middleboxes is identified based on the inport.}. 
Suppose we assign AS1 and AS2 traffic to Transcoder1, and AS3 and AS4
traffic to Transcoder2. Then CS2 matches on both the tag and the
prefix (more precisely, aggregated prefix of two base stations) to
forward AS1 and AS2 traffic to Transcoder1, and AS3 and AS4 traffic to
CS3. CS3 finally forwards AS3 and AS4 traffic to Transcoder2.

\subsection{Asymmetric Edge: Packet Classification}
To minimize data-plane state in the core, we want to classify packets
and associate them with tags as they enter the network.  The access
switch maintains microflow rules that, upon receiving a packet from a
UE, rewrite the IP address (to the location-dependent address) and
tags the packet with the policy tag.  Similarly, upon receiving
packets from the core network, the access switch rewrites the IP
address back to the value the UE expects.  The access switch learns
the appropriate rules from the controller.  For example, the access
switch could send the first packet of each flow to the controller, and
have the controller install the appropriate rule to direct the
remaining packets over a chosen policy path.  For better control-plane
scalability, the controller can provide a local agent at the base
station with appropriate classifiers for handling any traffic from
this UE, as discussed in more detail in Section~\ref{sec:control}.

Performing fine-grain packet classification is acceptable at the
access switch, due to the low link speeds and the relatively small
number of active flows.  However, gateway switches must handle orders
of magnitude more flows, and more flow-arrival events, so they should
not perform such fine-grained packet classification.  As such, we
adopt an asymmetric design where the gateway switches can easily
determine the appropriate IP address and tag to use, in one of two
ways:

\begin{figure}
\centering
\includegraphics[width=3.2in]{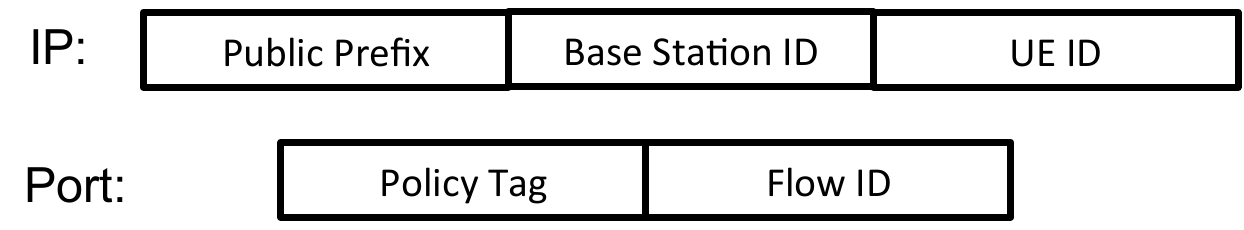}
\vspace*{-0.1in}
\caption{Embedding location and policy information in source IP address and source port number.
Thus the information can be implicitly piggybacked in return traffic.}
\vspace*{-0.1in}
\label{fig:encoding}
\end{figure}

\myitem{Embedding state in packet headers:} Rather than
\emph{encapsulating} packets, as is commonly done in data-center networks, we
can \emph{embed} the policy tag, base station ID, and UE ID in the
packet header.  This ensures that the return traffic carries these
fields.  For example, we could encode the state as part of the UE's IP
address (e.g., in IPv6), or a combination of the UE's IP address and
TCP/UDP port number (e.g., in IPv4) as shown in
Figure~\ref{fig:encoding}.  The access switch rewrites the source IP
address to the location-dependent IP address (i.e., the carrier's public prefix,
as well as the base station and UE IDs), and embeds the
policy tag as part of the source port.  UEs do not have many
active flows, leaving plenty of room for carrying the policy tag in
the port-number field.  With this embedding mechanism, our three
identifiers are implicitly ``piggybacked" in return traffic arriving
from the Internet\footnote{This approach raises some security and
  privacy challenges.  Malicious Internet hosts may spoof policy tags
  and congest network links or middleboxes, though these attacks
  can be blocked using conventional firewalls.
  In addition, a UE 's IP address changes upon moving to a new base
  station, making it easier for Internet servers to infer the UE's
  location.  Network address translation can reduce these concerns.}.
The gateway switch can simply make forwarding decisions based on the
destination IP address and port number of incoming packets.


\myitem{Caching state in gateway switches:} Instead of embedding state
in packet headers, the gateway switch can cache the state when
forwarding outgoing packets, and associate the state with the return traffic
arriving from the Internet. In this scheme, the gateway switch performs network address
translation, and caches the tag in the process.  In practice, network address
translation may be necessary anyway, if the cellular provider does not have a large
enough public IP address block to allocate a unique address for each
UE.  While NATing does introduce per-flow state, the gateway switch
does \emph{not} need to contact the controller or translate the UE's
address into the subscriber attributes.  While the gateway would need
a larger table than the other core switches, supporting microflow
rules does not require expensive TCAM or any sophisticated processing.

\subsection{Policy Consistency Under Mobility}
\label{sec:mobility}
Seamless handling of device mobility is a basic requirement for
cellular networks.  UEs move frequently from one base station to
another, and carriers have no control over when and where a UE moves.
In addition to minimizing packet loss and delay, carriers must ensure
that ongoing flows continue traversing the original sequence of
middleboxes (though not necessarily the same switches), while reaching
the UE at its new location.  Such \emph{policy consistency} is crucial
for traffic going through stateful middleboxes, like firewalls and
intrusion prevention systems.  However, new flows should traverse
middlebox instances closer to the UE's new location, for better
performance.  As such, SoftCell must differentiate between old and new
flows, and direct flows on the appropriate paths through the network.

\myitem{Differentiate between old and new flows:}
Incoming packets from old flows have a destination IP address
corresponding to the UE's old location, so these packets naturally
traverse the old sequence of middleboxes.  SoftCell merely needs to
direct these packets to the new base station, which then remaps the
old address to the UE's permanent address.  During the transition, the
controller does not assign the old location-dependent address to
any new UEs.
%
For the traffic sent \emph{from} the UE, the old access switch has a
complete list of microflow rules for the active flows.  Copying these
rules to the new access switch ensures that packets in these
flows continue to be mapped to the old IP address, to avoid a disruption in
service.  Each UE has a relatively small number of active connections
(say, 10), limiting the overhead of copying the rules.  To minimize
hand-off latency, the SoftCell controller could copy these rules in
advance, as soon as a UE moves near a new base station.

\begin{figure}
\centering
\includegraphics[width=2.5in]{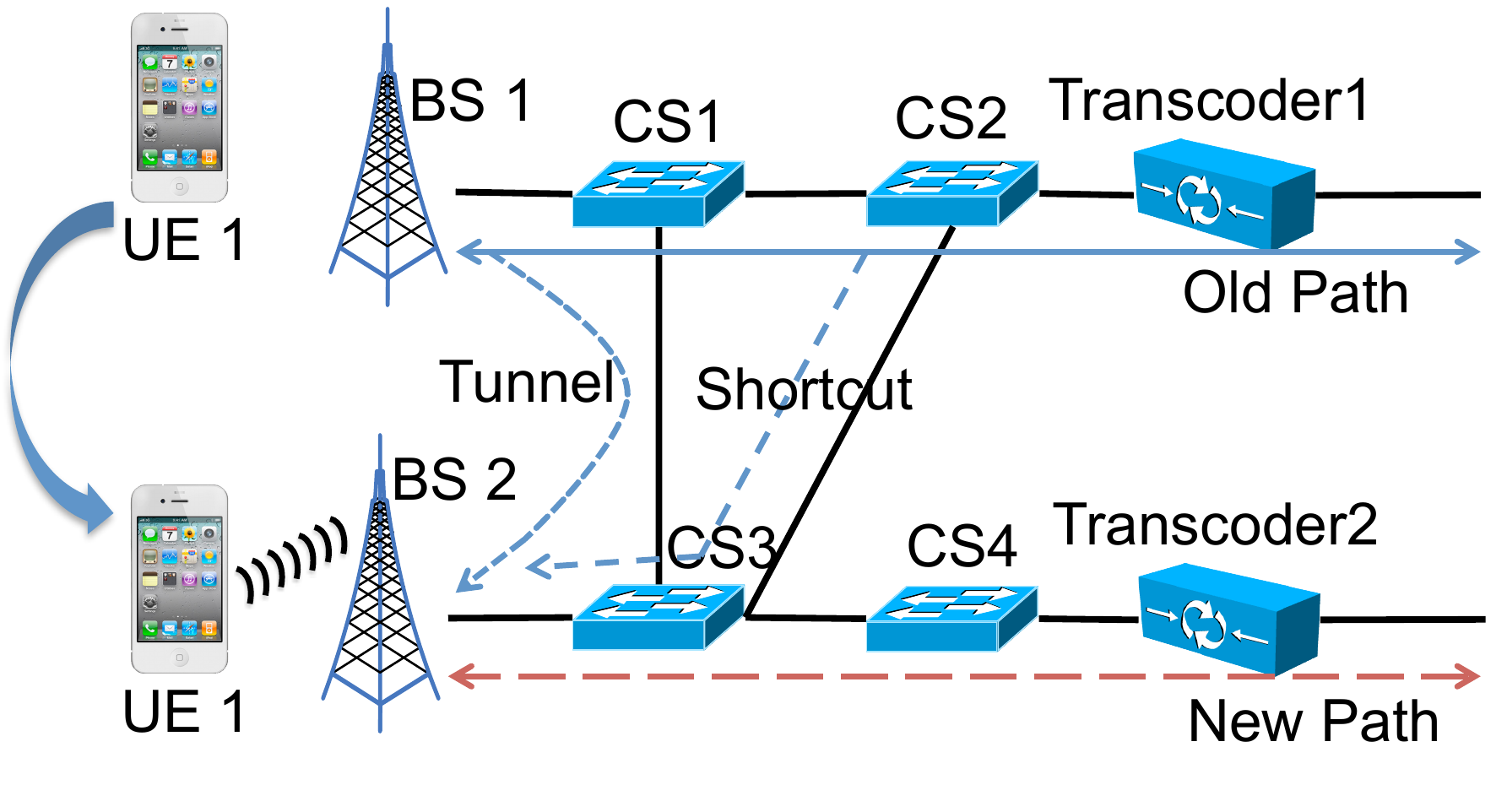}
\vspace*{-0.15in}
\caption{Tunnels and shortcuts for old flows}
\vspace*{-0.1in}
\label{fig:mobility}
\end{figure}

\myitem{Efficiently reroute the old flows:}
To handle ongoing connections during mobility events, SoftCell
maintains long-lived \emph{tunnels} between nearby base stations, as
shown in Figure~\ref{fig:mobility}.  These tunnels can carry traffic
for \emph{any} UEs that have moved from one base station to another.  This
``triangle routing'' ensures policy consistency and minimizes packet
loss, at the expense of higher latency and bandwidth consumption.  The
many short-lived connections would not experience any significant
performance penalty.
To handle long-lived connections more efficiently, the controller can
establish temporary \emph{shortcut} paths for directing traffic
between the new base station and the old policy path, as shown in
Figure~\ref{fig:mobility}.  The controller can learn the list of
active microflows from the access switch at the old base station, and
install rules in the core switches to direct incoming packets over the
shortcut paths.  A single UE may need multiple shortcuts, since
different traffic may go through different middleboxes\footnote{No
  short-cut paths are needed in the common case when a UE moves to another base station
  in the same \emph{cluster}, since these base stations connect to the same
  core switch.  In this case, simply adding the microflow rules at
  this core switch is sufficient.}.  As such, these shortcut paths are
created when a UE moves, and removed when a soft timeout
expires---indicating that the old flow has ended.


\subsection{Rule Minimization in Core Switches}
\label{sec:algo}

We have shown that we can reduce the number of core switch rules by relying on
multi-dimensional aggregation. 
We now present an online algorithm that performs policy path implementation in
real time on a per policy path basis. For ease of description, we first describe
our path implementation algorithm assuming the policy path is a simple path, as
shown in Algorithm \ref{alg:path}. We then discuss how to deal with loops.   
 
\myitem{Simple tag reuse rules:}
To reduce the amount of switch rules, we want to maximize the reuse of
existing rules which match policy tags and base station IDs. Our first
step is to pick a tag that is already used in switches where the new policy path
includes. To ensure correctness, we impose the constraint that different policy
paths originated from the same destination access switch to have different
tags. Otherwise, we would not be able to distinguish among different policy
paths from the same base station.  
As shown in Algorithm~\ref{alg:path}, we enumerate the candidate tags
and choose the tag that results the minimal number of new rules we need to install
(line 1-8). 
It is possible that the candidate tags are an empty set, in which case we will choose a random
unused tag (line 8).

\myitem{Safe aggregation:}
When we iterate over each switch for a given tag $t$ along
the policy path, for each switch, we calculate how many rules we need to
install with the candidate tag and base station prefix (line 4). This is done in
method $sw_1.getNewRule(t, prefix, sw_2)$. This method performs~\emph{safe aggregation}.
In particular, the method looks at all switch rules with tag
$t$ with an action that forwards to the same next-hop switch $sw_2$. It will try to
aggregate the base station prefixes of the rules. By safe aggregation, we mean
that resulting prefix of the aggregate rule contains the exact number of
component prefixes. For example, if there are three /16 base station prefixes,
we can not aggregate them into a /14 prefix. On the other hand, if we have all
four component prefixes, we can aggregate them into the /14 prefix.
After picking the tag $tag^*$ to use, the algorithm installs the path with the
prefix and the tag. We only need to install rules to switches where we cannot utilize
existing rules to reach the correct next hop (line 10-15). We can install
aggregate rules if the aggregation is safe (line 13). Otherwise, we just install
this rule (line 15). Note that safe aggregation is done \emph{atomically} to prevent
inconsistencies introduced by race condition.

\myitem{Dealing with loops:}
Ideally, we should only compute and install loop-free paths. 
However, due to the flexibility of service policies and placements of middleboxes,
loops are sometimes unavoidable. For instance, in Figure \ref{fig:high-level}, there is 
no way to avoid a loop in the path if a service policy clause requires outbound
video traffic to go through a firewall before a video transcoder.
A loop that enters a switch twice but from different links can be
easily differentiated by input ports. However, a loop that 
enters a switch twice from the \emph{same} link is more difficult to handle. In such
a case, we use additional tags to help switches make forwarding decisions. More
specifically, we break a loop into two segments; each segment uses one tag for
forwarding. At the switch that connects these two segments, we install a rule to
``swap'' these two tags. This approach can be generalized to support nested loops. 

\newcommand{\argmin}{\operatornamewithlimits{argmin}}
\begin{algorithm}[t]
\caption{Install A New Policy Path }
\footnotesize
\textbf{Input:} \begin{itemize}[noitemsep,nolistsep]
    \item[--] $path$: the policy path to install
    \item[--] $prefix$: the IP prefix of the base station
    \item[--] $candTag$: the set of candidate tags for the base station 
    \item[--] $usedTag$: the set of tags used by the base station 
\end{itemize}
\textbf{Output:} $switch~rules$ and a $tag$ for this policy path

\begin{algorithmic}[1]
\Statex \textbf{Step 1: Choose a tag to minimize new rules}
\For{$t$ in $candTag$}
        \State $newRule[t] = 0$ \Comment new rules needed if tag $t$ is used
	\For{$(sw_1, sw_2)$ in $path$}
	    \State $newRule[t]+=sw_1.getNewRule(t, prefix, sw_2)$
	\EndFor
\EndFor

\If{$candTag \, != \emptyset$}
        \State $tag^* = \underset{t}{\operatorname{arg\,min}} \{newRule[t]\} $
\Else
	\State $tag^* = random \{t | t \notin usedTags\}$
\EndIf

\State $usedTag = \{tag^*\} \cup usedTag$

\Statex \textbf{Step 2: Install the path with the prefix and tag}
\For{$(sw_1,sw_2)$ in $path$}
	\If{$sw_1.getNextHop(tag^*, prefix) != sw_2$}
		\If{$sw.canAggregate(tag^*, prefix, sw_2)$}
			\State $sw.aggregateRule(tag^*, prefix, sw_2)$
		\Else
			\State $sw.installRule(tag^*, prefix, sw_2)$
		\EndIf
	\EndIf
\EndFor

\end{algorithmic}
\label{alg:path}
\end{algorithm}

\section{Scalable Control Plane}
\label{sec:control}


Sending the first packet of every flow to the central controller would
introduce a high overhead.  Instead, a local agent at each base
station offloads part of the control-plane functionality.  In this
section, we first present the design of the local agent and then
describe how the control plane handles network dynamics.

\begin{table*}[t]
\centering 
{\small
\begin{tabular}{|l|c|c|c|c|c|}
\hline
 & \multirow{2}{*}{\bf Access Switch}  & \multirow{2}{*}{\bf Core Switch} & {\bf Controller} & \multicolumn{2}{|c|}{\bf Controller w/ Local Agent} \\
\cline{5-6}
& & & {\bf w/o Local Agent} & {\bf Central Controller} & {\bf Local Agent} \\
\hline
{\bf UE Arrival} & Yes & No & Yes & Yes & Yes \\
\hline
{\bf Flow Arrival} & Yes & Sometimes & Yes & Sometimes & Yes \\
\hline
{\bf UE Handoff } & Yes & Yes(Relevant) & Yes & Yes & Yes\\
\hline
{\bf Topology Change} & No & Yes(Relevant) & Yes & Yes & No \\
\hline
{\bf Dynamic Policy} & No & Yes(Relevant) & Yes & Yes & No \\
\hline
\end{tabular}
}
\caption{How network events affect data plane and control plane.
\emph{Yes} means the switch/controller is involved in the event, \emph{No} means not involved,
\emph{Sometimes} in \emph{Flow Arrival} means only involved when the policy path has not been installed,
and \emph{Relevant} means only relevant switches (a small number of the whole) are involved.
Central controller offloads most \emph{Flow Arrival} events to local agents.
}
\label{tab:event}
\end{table*}

\subsection{SoftCell Local Agent}
\label{sec:agent}
Each base station runs a local software agent equipped with the
computing power to conduct various management tasks, including radio
resource allocation for UEs. The local agent caches a list of
\emph{packet classifiers} for each UE at the behest of the central
controller. The packet classifiers are a~\emph{UE-specific}
instantiation of the service policy that matches on header fields in
the packet and identifies the appropriate policy tag, if a policy path
already exists.  When the UE arrives at the base station, the
controller computes the packet classifiers based on the service
policy, the UE's subscriber attributes, and the current policy tags.
When the UE starts a new flow, the local agent consults these
classifiers to determine the right policy tag for these packets, and
installs a microflow rule in the access switch, similar to the
``clone'' function in DevoFlow~\cite{devoflow}.  The local agent only
contacts the controller if no policy tag exists for this
traffic---that is, if the packet is the first traffic at this base
station, across all UEs, that need a particular policy path.

Let's use an example to illustrate this. Suppose UE7 arrives at base
station 1 with prefix 10.0.0.0/16. The local agent first assigns a UE
ID \emph{10} to the UE. Now UE7 is associated with the
location-dependent address 10.0.0.10. The local agents also contacts
the controller to fetch a list of packet classifiers for this
UE. Suppose the list includes two packet classifiers:

\begin{tabular}{ll}
1. match:dst\_port=80, & action:tag=2\\
2. match:dst\_port=22, & action:send-to-controller
\end{tabular}

\noindent
When a packet with destination port 80 from UE7 arrives, the access
switch find any existing microflow rule, and directs the packet to the
local agent.  The local agent determines that the traffic matches the
first packet classifier. Since the policy path already exists, the
local agent simply installs a microflow rule in the access switch
which (i) rewrites the UE IP address to 10.0.0.10 and (ii) pushes
``tag=2" to the source port number, without contacting the central
controller.  Suppose another packet arrives from UE7 with destination
port 22. This flow matches the second packet classifier and the action
is ``send to controller". This means the policy path to base station 1
has \emph{not} been installed yet.  The local agent sends a request to
the central controller to install a new policy path and returns the
policy tag. Then, the local agent can update the packet classifier and
install a microflow rule for the packets of this flow.

In this way, local agents cache UE-specific packet classifiers and
process most microflows locally, significantly reducing the load on
the controller.

\subsection{Handling Network Dynamics}
Next, we discuss how the control plane deals with network dynamics, as
summarized in Table~\ref{tab:event}. We already discussed UE and
flow arrival in Section~\ref{sec:agent}, and UE handoff in
Section~\ref{sec:mobility} respectively. Here, we just briefly discuss
topology change and dynamic policy.

\myitem{Topology change:} When the network topology changes (due to a
link, switch, or middlebox failure), the controller calculates and
installs new paths for affected policies. In some cases, like a
stateful middlebox crash without any state saved, in-progress flows
may experience significant packet loss or even have to terminate. The
topology change is handled by the controller and only affects relevant
switches. There is no need to update all access switches to change
their flow table.

\myitem{Dynamic policy:}
In addition to static policy, SoftCell supports dynamic policy
which change the policy path during the lifetime of a flow.
For example, when the air interface of a base station is congested,
the service policy may require video traffic to go through a
transcoder. In this case, the controller must install a new path for
video traffic.  The central controller updates the policy paths in the
core network, based on the policy requirements, without changing the
policy tags.

\section{Extensible Controller Design}
\begin{figure}
\centering
\includegraphics[width=3.4in]{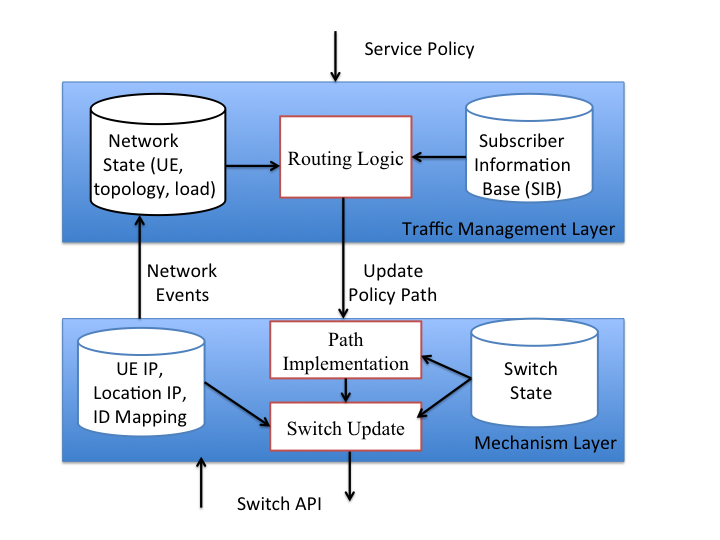}
\vspace*{-0.35in}
\caption{SoftCell controller}
\vspace*{-0.1in}
\label{fig:controller}
\end{figure}

In addition to supporting service policies, carriers need to manage
their network and middlebox resources, to minimize latency and balance
load.  Our controller design cleanly separates traffic management from
the low-level mechanisms for installing rules and minimizing
data-plane state, as shown in Figure~\ref{fig:controller}.

\myitem{Traffic-management layer:} The traffic-management layer
computes policy paths through the switches and middleboxes, to satisfy
both the service policy and traffic-management goals.  This layer
determines the service attributes for a UE from the Subscriber
Information Base (SIB), and consults the service policy to compute
policy paths that traverse the appropriate middleboxes and optimize
traffic-management objectives.

\myitem{Mechanism layer:} The mechanism layer realizes the policy
paths by installing rules in the underlying switches, using the
techniques proposed in the previous two sections.  This layer hides
all the details of location-dependent addresses, the encoding of
policy tags, the path implementation algorithm, and assuring path
consistency during mobility.  The mechanism layer could also poll
traffic counters in the switches and aggregate them to the level of
policy tags to enable the traffic-management layer to operate on a
coarser-grain view of the traffic.

A modular controller design allows each layer to evolve independently,
to adopt new innovations in how to manage traffic and data-plane
state, respectively.

\section{Performance Evaluation}
\label{sec:impl}
In this section, we demonstrate the scalability and performance of our SoftCell
architecture. First, we measure the workload that SoftCell would face in a typical
cellular core network by analyzing a trace from a large LTE network. We then show
that SoftCell is able to sustain several times of this workload by performing micro
benchmark. Finally, we show that SoftCell can handle thousands of service policy
clauses on commodity switches through large-scale simulations.

\begin{figure*}[ht]
\centering
\subfigure[Events in the whole network]{
\label{fig:meas-whole}
\includegraphics[width=2.2in]{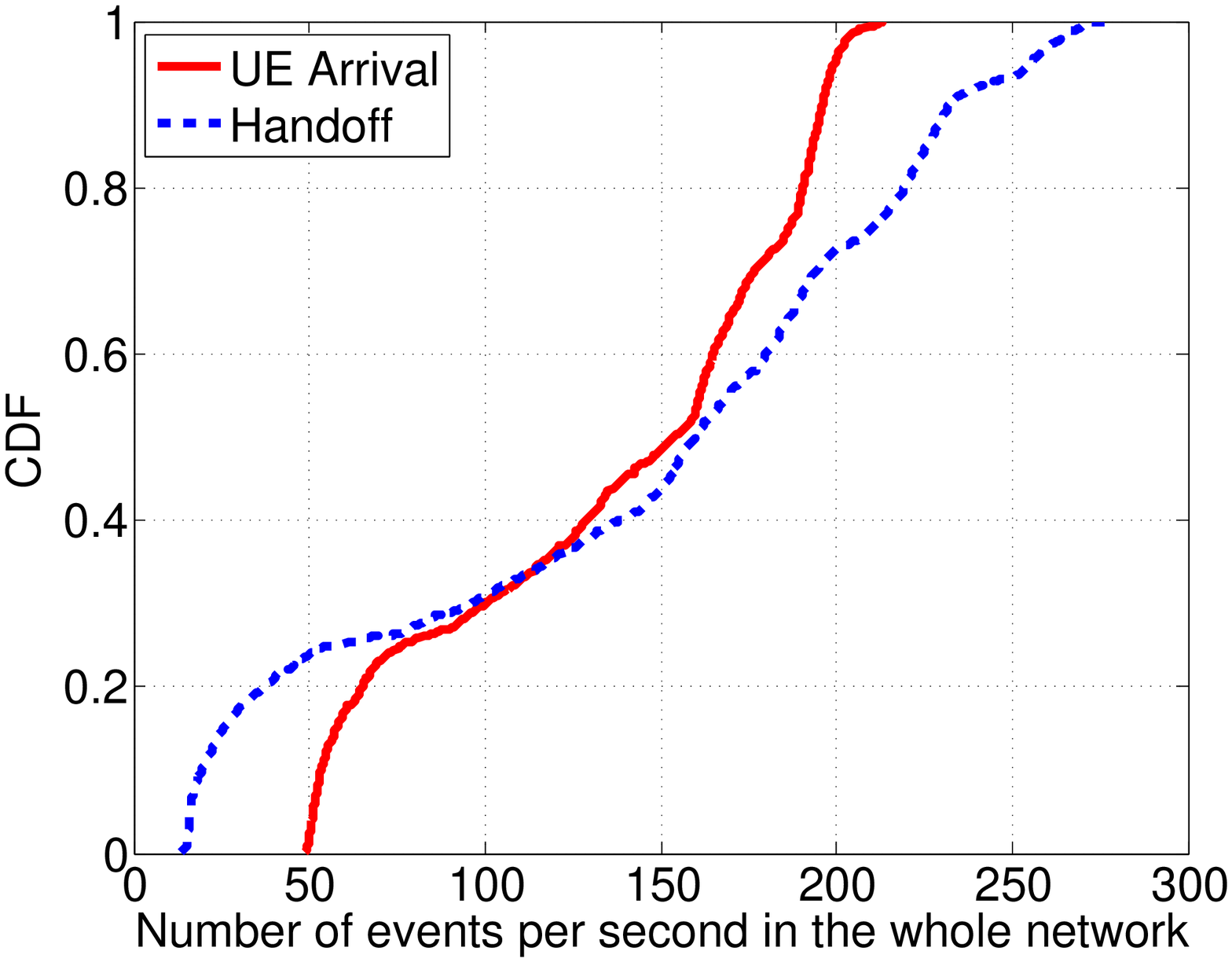}}
\subfigure[Active UEs per base station]{
\label{fig:meas-active-ue}
\includegraphics[width=2.2in]{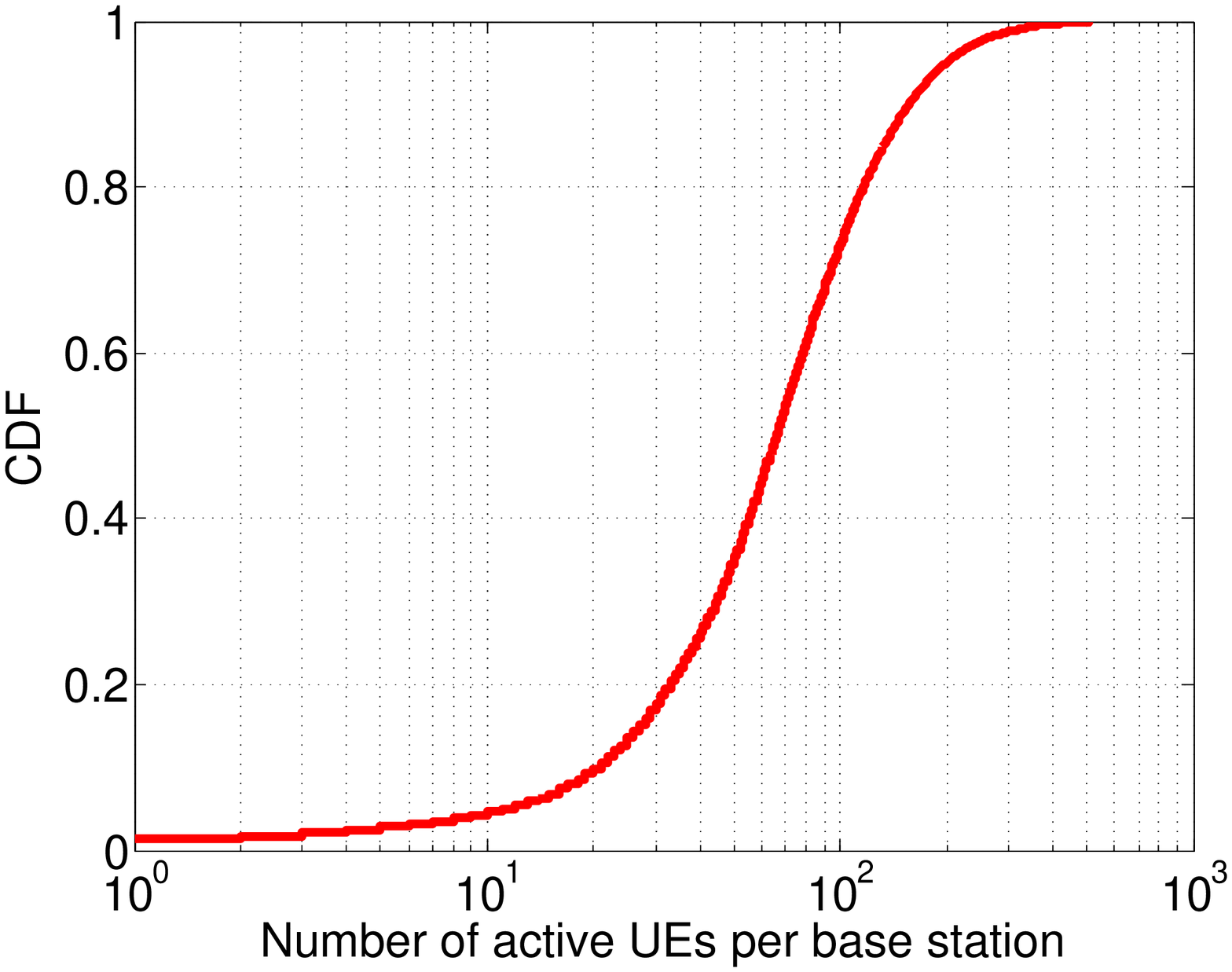}}
\subfigure[Radio bearer
arrivals per base station]{
\label{fig:meas-bearer-arrival}
\includegraphics[width=2.2in]{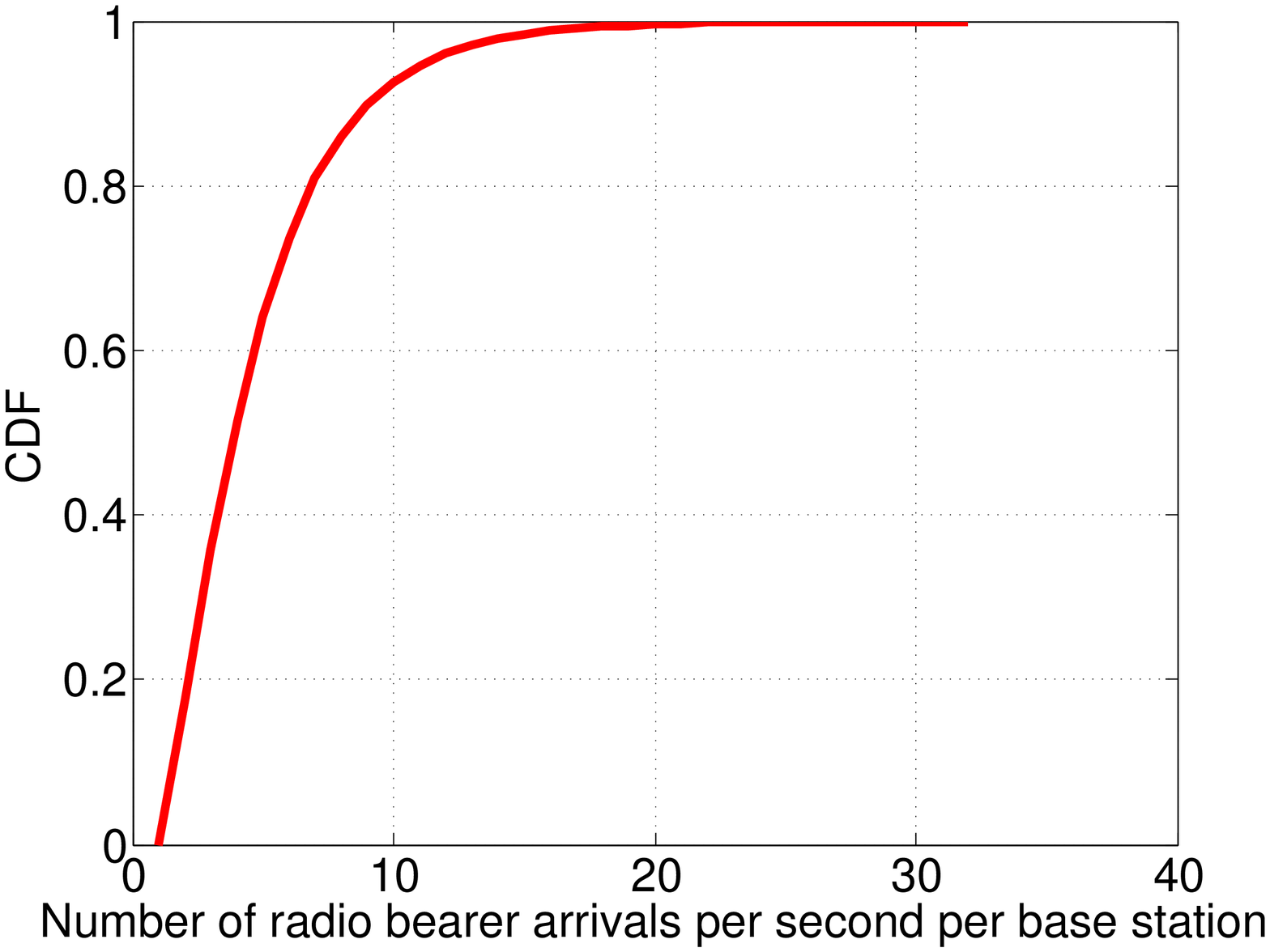}}
\vspace*{-0.15in}
\caption{Measurement Results of a LTE network}
\vspace*{-0.1in}
\label{fig:meas}
\end{figure*}

\subsection{LTE Workload Characteristics} 
As a first step towards SoftCell deployment, we measured the workload of a real
cellular network to understand the practical performance requirements of the
controller.

\myitem{Dataset Description:} We collected about 1TB traces from a large ISP's LTE
network during one week in January 2013.  The dataset covers a large metropolitan
area with roughly 1500 base stations and 1 million mobile devices (including mobile
phones and tablets).  The trace is bearer-level and includes various events such as
radio bearer creation, UE arrival to the network, UE handoff between base stations,
etc.  A radio bearer is a communication channel between a UE and its associated base
station with a defined Quality of Service (QoS) class.  When a flow arrives and there
is an existing radio bearer with the same QoS class, the flow will use the existing
radio bearer.  Since radio bearers timeout in a few seconds, it is possible that a
long flow may trigger several radio bearer creation and deletion. Since we do not
have flow-level information, we use radio bearers as an estimation of flow activity.
We present measurement results for a typical week day.

\myitem{Network wide characteristics:} Figure \ref{fig:meas-whole} shows the CDF of
UE arrival events and handoffs in the whole network.  A UE arrival event means a new
UE first attaches to the network, e.g., after a UE is powered on.  When a UE arrives
at the network, the central controller fetches the UE attributes from the SIB and
send the UE's packet classifiers to the local agent.  A UE handoff event means a UE
transfers from one base station to another.  Upon handoff, the controller has to copy
state from the old access switch to the new access switch with the help of local
agents, and set up shortcuts for long flows.  We do not account for UE handoffs
between cells of the same base station as they do not cause forwarding changes. From
the figure, we can see that the 99.999 percentile of UE arrival and handoff events
per second are 214 and 280, respectively.
While each of these events requires the central controller to contact local agents or
update core switches, it is not a problem as today's commodity servers and software
switches can easily handle tens of thousands of these events per second. Actually,
even if we account for the exponential growth of mobile data~\cite{MobileData18} (18
times in the next five years), the workload can still be easily handled by commodity
servers and software switches.

\myitem{Load on each base station:} Figure \ref{fig:meas-active-ue} shows the CDF of
active UEs per base station. We see that a typical base station handles hundreds of
active UEs simultaneously, with a 99.999 percentile of 514.  Figure
\ref{fig:meas-bearer-arrival} depicts the radio bearer arrival rate at each base
station. The number is relatively small, i.e. only 34 for the 99.999 percentile. As
one radio bearer typically carries a handful of concurrent
flows~\cite{rahmati,zhang12}, we expect the actual flow arrival rate to be around
several hundred flows per second. These results imply that the local agent has to
keep state for several hundred of UEs and process a maximum of tens of thousands new
flows per second. However, as most of the time policy paths would have already been
installed in the network, new flow requests only require the local agent to install
packet classification rules at the access switch. Again, tens of thousands of these
events per second can be easily handled by today's software switches.

\subsection{Controller Micro Benchmark}
We have implemented a SoftCell control plane prototype on top of the popular
Floodlight~\cite{floodlight} OpenFlow controller.  The prototype implements both
SoftCell central controller and SoftCell local agent.  To perform topology discovery,
we rely on the ``TopologyService" module provided by Floodlight.  Since there is no
support for middlebox discovery in Floodlight, we encode middlebox placement and
specification in a configuration file that is provided to the controller. The local
agent fetches packet classifiers from the global controller upon every new UE
arrival, then it uses the packet classifiers to handle all the following flows from
the new UE. The communication between a local agent and the global controller is
implemented with the Floodlight REST API.

In the following, we perform micro benchmark on the prototype, then we compare the
results with the measurement results obtained earlier to demonstrate the ability of
our controller to sustain the workload. We benchmark the prototype using
Cbench~\cite{cbench}. Cbench emulates a number of switches, generates packet-in
events to the tested controller, and counts how many events the controller processes
per second (throughput). Each test server has an intel XEON W5580 processor with 8
cores and 6GB of RAM.

\myitem{Central controller performance:} First, we evaluate the throughput of the
controller. Recall that the controller has to send packet classifiers to local agents
when a UE attaches or moves to a base station. We use Cbench to emulates 1000
switches and let these switches keep sending packet-in events to the controller. From
the controller viewpoint, these packet-in events correspond to packet classifier
requests coming from 1000 local agents. The controller then replies to these requests
with packet classifiers as fast as it can. 

Our results show that the controller can process 2.2 million of requests per second
with 15 threads. These results clearly demonstrate that the SoftCell controller can
sustain the load of a large LTE network as, from our measurement study, we know that
only hundreds to thousands of such events are encountered per second. Also, these
results are similar to the ones reported by~\cite{controller-perf}, the difference
being mainly due to different settings and platforms.

\myitem{Local agent performance:} Second, we evaluate the throughput of the local
agent. Recall that the local agent needs to fetch packet classifiers from the
controller when processing events. The throughput of the local agent therefore
depends on how frequently it needs to contact the controller which itself depends on
the cache hit ratio. Table~\ref{tab:micro-cache-hit} shows the evolution of the local
agent throughput in function of the cache hit ratio. A cache hit ratio of 80\% means
that the local agent can handle 80\% of events locally and need to contact the
controller for the remaining 20\% of the events. To measure the throughput, we use
Cbench to emulate the access switch connected to the local agent and let it keep
sending packet-in events to the local agent. Upon the reception of a packet-in event,
the local agent performs a lookup in its cache and contact the controller upon cache
misses. The local agent and the controller run on two separate servers connected by
the department LAN. 

Again, the local agent throughput is sufficient to handle the number of new flows
measured at a base station (a few to tens of thousands per second). Indeed, even in
the worst case where the local agent has to contact the controller for every event,
it is still able to handle 1.8K events per second. Not to mention that we can also
optimize the cache hit ratio, e.g., by prefetching packet classifiers from the
controller.

\begin{table}[t]
\centering {\small
\begin{tabular}{|l|c|c|c|c|c|c|}
\hline
Cache Hit
Ratio & 0\% & 20\% & 40\% & 60\% & 80\% & 100\% \\
\hline
Throughput & 1.8K & 2.3K &
3.0K & 4.5K & 8.6K & 505.8K \\
\hline \end{tabular} }
\vspace*{-0.15in}
\caption{Effect of cache hit ratio on local agent throughput}
\vspace*{-0.15in}
\label{tab:micro-cache-hit}
\end{table}

\begin{figure*}[t]
\centering
\subfigure[Effect of the number of service policy clauses]{
\label{fig:simul-number}
\includegraphics[width=2.2in]{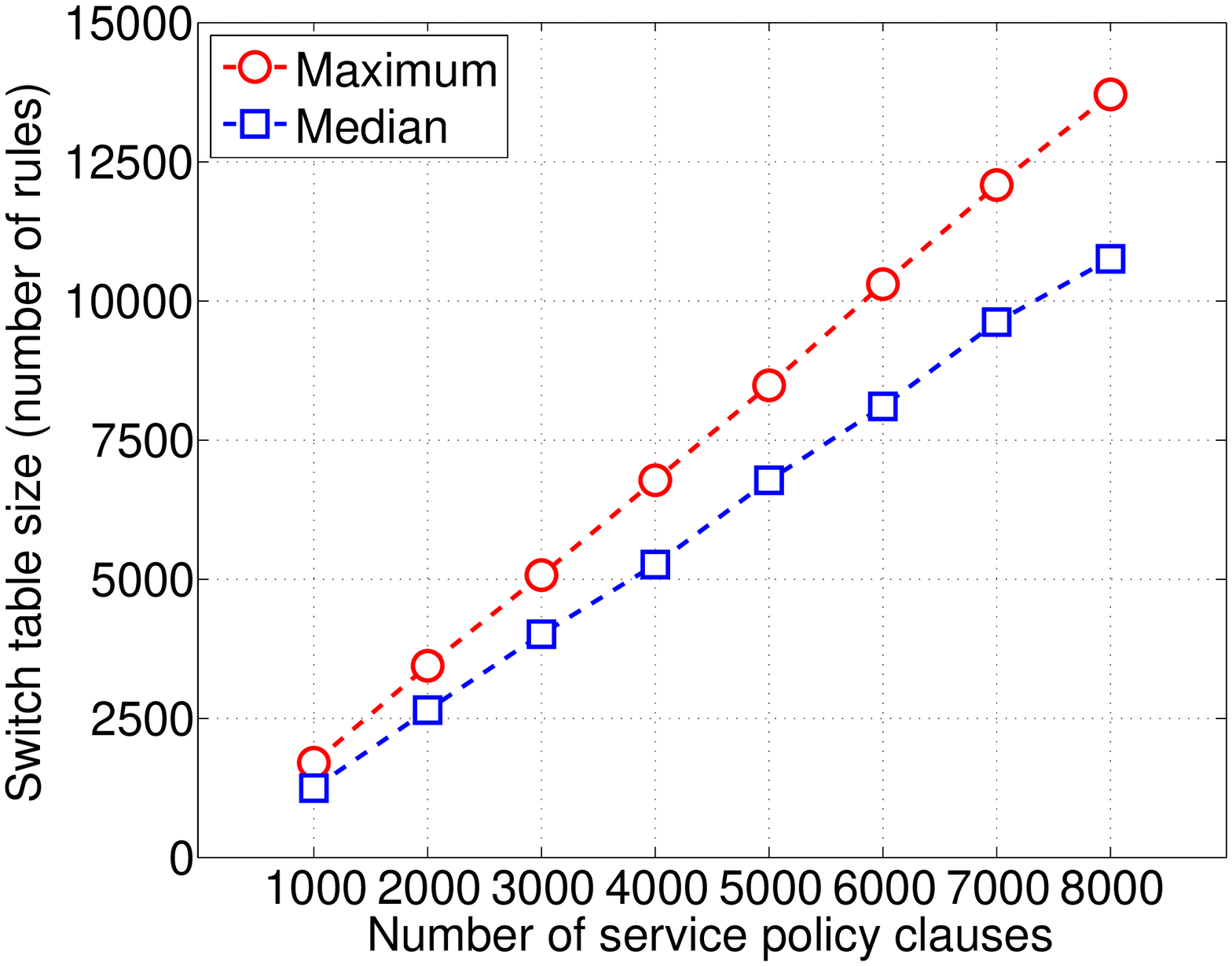}}
\subfigure[Effect of service policy clause length]{
\label{fig:simul-length}
\includegraphics[width=2.2in]{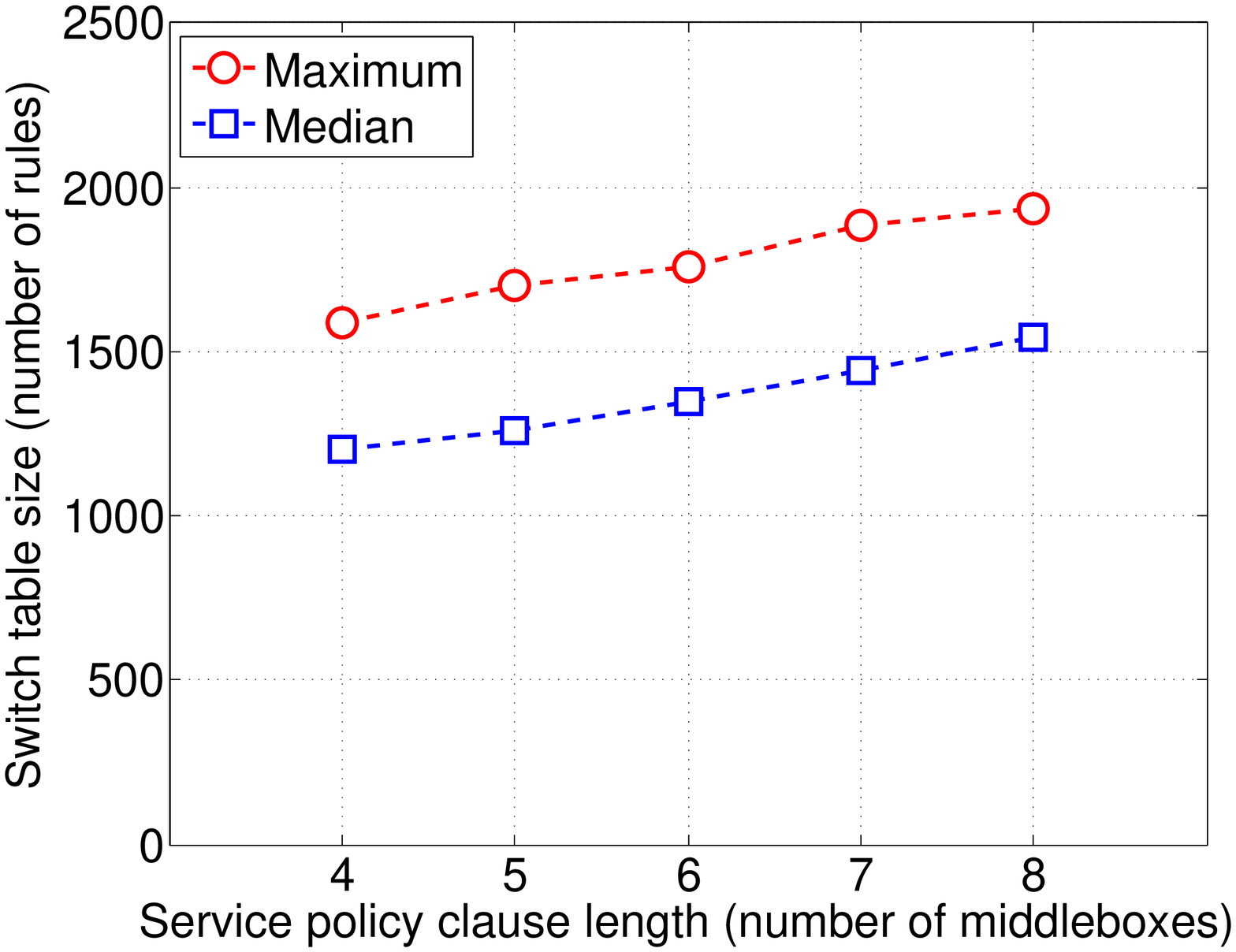}}
\subfigure[Effect of network size]{
\label{fig:simul-size}
\includegraphics[width=2.2in]{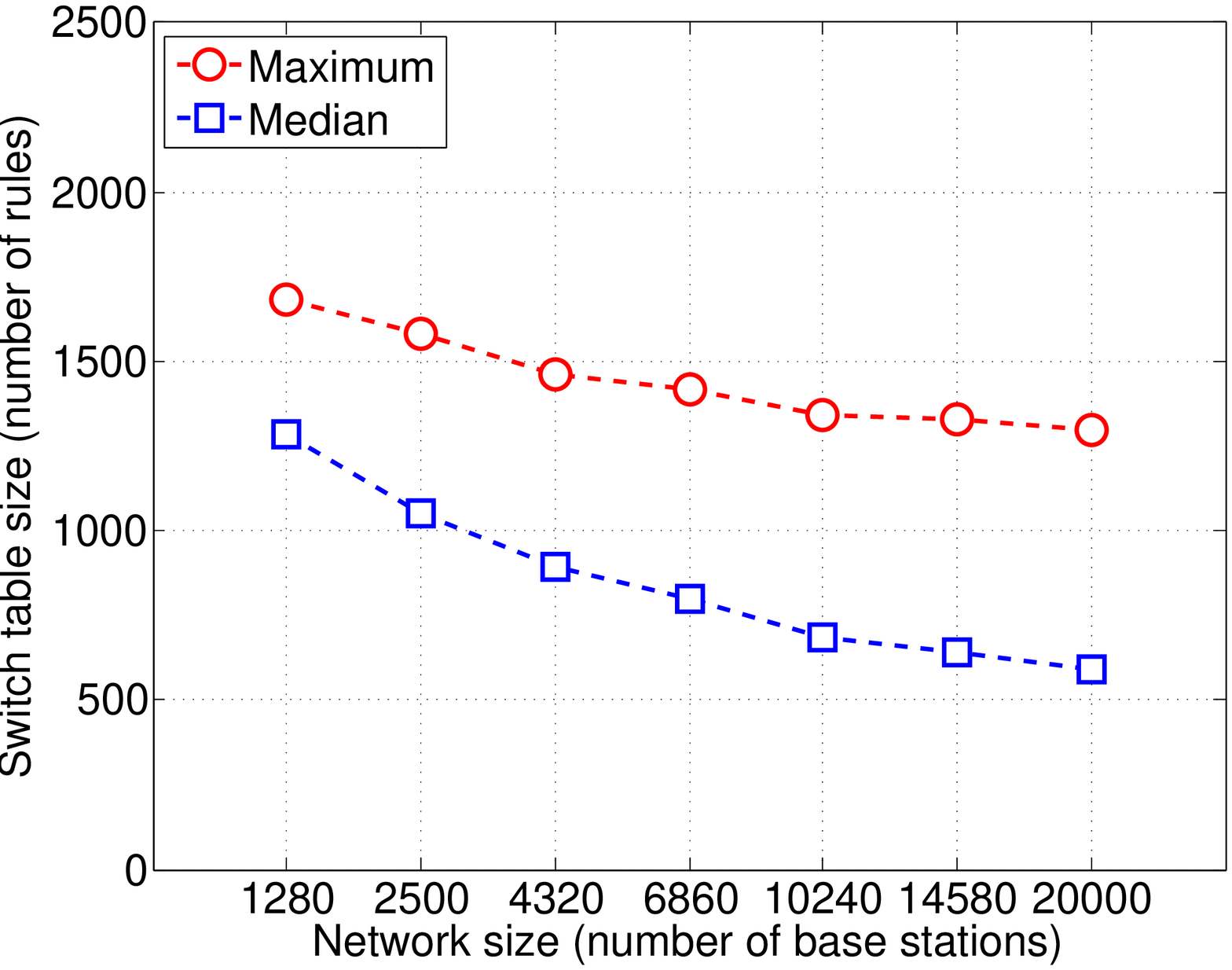}}
\vspace*{-0.15in}
\caption{Large-scale simulation result.  With multi-dimensional
aggregation, SoftCell data plane is able to support thousands of service policy
clauses on commodity switches.}
\vspace*{-0.1in}
\label{fig:simul}
\end{figure*}

\subsection{Large-Scale Simulations} We now demonstrate the scalability of SoftCell
data plane through large scale simulations. In particular, we show that SoftCell only
requires a few thousand TCAM entries to support thousands of service policy clauses
for thousands of base stations.

\myitem{Methodology:} We generate hierarchical topology composed following the
description of cellular core networks in~\cite{celltopology, bscluster}. Each
topology is composed of three layers: \emph{access}, \emph{aggregation} and
\emph{core}. The access layer is composed of a cluster of 10 base stations
interconnected in a ring fashion. Among these 10 base stations, two of them is
connected to the aggregation layer~\cite{bscluster}. The aggregation layer is
composed of $k$ \emph{pods}, each of which is composed of $k$ switches connected in
full-mesh. In each pod, $k/2$ switches are connected to $k/2$ base station clusters.
The remaining $k/2$ switches are connected to $k/2$ switches residing in the core
layer. The core layer is itself composed of $k^2$ switches connected in full-mesh.
Each core switch is furthermore connected to a gateway switch. The whole topology is
composed of $10k^3/4$ base stations. For example, $k=8$ (resp. $k=20$) gives a
network with $1280$ (resp. $20000$) base stations. For each topology, we assume that
they are $k$ different types of middleboxes. We randomly connect one instance of each
type in each pod composing the aggregation layer and two instances of each type in
the core layer. On top of this topology, we generate $n$ policy paths for \emph{each}
base station to the gateway switch. A policy path traverses $m$ randomly chosen
middlebox instances. Finally, we measure the number of rules in each switch flow
table. In the base case, we consider $n=1000$, $m=5$ and $k=8$. We vary $k$, $n$ and
$m$ to show how the switch state is affected by the number of service policy clauses,
the policy length and the network size, respectively.

\myitem{Effect of number of service policy clauses:} Figure~\ref{fig:simul-number}
shows the maximum and median size of the switch forwarding table with respect to the
number of service policy clauses. We can see that switch table size increases
linearly with the number of service policy clauses with a small slope (less than 2).
In particular, to support $1000$ service policy clauses (1.28 million policy paths!),
switches store a median of $1214$ rules and a maximum of $1697$ rules. Even to
support $8000$ service policy clauses, the maximum table size is only $13682$.
Observe that in practice ISPs may only need tens or hundreds of service policy
clauses, meaning that SoftCell can be easily implemented on commodity switches. The
good performance of SoftCell data plane is a direct consequence of its
multi-dimensional aggregation (see Section~\ref{sec:data}) capability. Indeed, even
if a service policy clause instantiates a policy path to each base station, the
corresponding forwarding entries can be aggregated by prefix in the core layer
provided that they traverse the same middlebox instance like CS1 in
Figure~\ref{fig:ex}(c). Similarly, in the aggregation layer, the forwarding entries
corresponding to paths traversing the same middlebox instance in a pod can be
aggregated by prefix like CS2 and CS3 in Figure~\ref{fig:ex}(c). As such, to install
a new service policy clause, each switch only installs a handful of new rules in
average. 

\myitem{Effect of service policy clause length:} Figure~\ref{fig:simul-length} shows
the switch table size with respect to the policy length. When the maximum service
policy clause length is $8$, the maximum switch table size is $1934$. As before, we
see that switch table size increases linearly with the length of the service policy
clause with a small slope. Indeed, when a service policy clause is longer, the policy
paths traverse more middleboxes and require more rules for forwarding. However, most
affected switches on the path only need one additional rule to match on the tag; only
a few switches are connected to multiple middleboxes and therefore need to dispatch
traffic to multiple middlebox instances. Thus the switch table size increases slowly
across all policy clauses. Again, observe that a service policy clause length of 8
(traversing 8 middleboxes) is an aggressive number, while in practice 4 or 5 is
sufficient.

\myitem{Effect of network size:} Figure~\ref{fig:simul-size} shows the switch table
size with respect to the network size. We see the table size decreases as the network
grows. It is true that with more base stations, we have to install more policy paths
for the same service policy clause, thus need more rules. But remember that we can
do aggregation on policy tags and base station prefixes, and when the network
increases, we have more switches. The increase of rules is small due to aggregation
and all rules are distributed over the more switches.
This leads to the result that when the network grows,
switches maintain smaller tables for the same number of service policy clauses.
\afteritem

In summary, SoftCell can support thousands of service policy clauses in a network of
thousands of base stations with a few thousand TCAM entries, which can be easily
achieved by commodity switches. The gain essentially comes from the ability to
selectively match on multiple dimensions.


\vspace{-0.1in}
\section{Discussion}
\label{sec:disc}

\myitem{Traffic initiated from the Internet:}
Although most traffic in cellular networks today are initiated from UEs,
some carriers~\cite{attIP} also offer various public IP address
options. 
When a gateway switch receives packets destined to these special public IP
addresses, the gateway will act like an access switch.
It will install packet classifiers that translate the public IP addresses
and the port numbers (with which UEs provide service to the Internet)
to LocIPs and policy tags. Note that these packet classifiers are not microflow
rules and don't require communication with the central controller for every
microflow. They are coarse grained (match on the UE public IPs and port
numbers) and can be installed once. 


\myitem{Asymmetric Internet routing:}
%
For ease of description, we have assumed that flows leaving a gateway switch
return to the same gateway switch. However, Internet routing is not guaranteed
to be symmetric. If gateway switches are not border routers peering with other
autonomous systems, border routers can be configured to route return traffic to
the same gateway switch. Alternatively, the controller can install corresponding
switch rules for return traffic in all possible gateway switches (mostly a small
fraction of the total number of gateway switches).

\myitem{Exact rule matching switches:}
Our design and evaluation of SoftCell has assumed that switches can do
prefix-matching on IP address and port number. To extend
SoftCell to handle exact rule matching switches, there are two cases. In the
case we embed state in packet headers,  SoftCell requires a special gateway
switch that can copy location IP prefix and tag information from IP headers to
fixed fields (if exists and not used for other purpose or append a header like
MPLS) these switches can match. This is a very simple function (copy some bits
from some fields to other fields) that doesn't need to store any state for
execution and can be implemented in hardware with line speeds. 
In the case of caching state at gateway switches, our wild card rule can be in
the control plane of the gateway switches. We can use mechanism like
Devoflow~\cite{devoflow} to install micro flow rules on demand.

\myitem{On-path middleboxes:}
%
The only problem with on-path middleboxes is that it is unavoidable to traverse
them in some cases. If service policy specifies that certain flows can not
traverse certain middleboxes (which we have not considered in our service
policy), then our path computation has to avoid these middleboxes. In case no
feasible path exists, the policy path request will be denied. 

\myitem{Radio resource control state tracking, paging and roaming:}
Base stations keep track of UE Radio Resource Control State (RRC) state and
the SoftCell controller keeps track of the current location area of a UE.
Our handling of RRC state tracking and paging is in principle the same as current
LTE. Roaming traffic are handled the same way as native traffic albeit with
different service policy. How to obtain roaming subscriber information for
authentication, and how to do billing etc are coordinated among controllers of
carriers. We do not discuss the details in this paper.

\vspace{-0.05in}
\section{Related Work}
\label{sec:related}
Our quest is to build a scalable architecture to support fine-grained policies
for mobile devices in cellular core networks. SoftCell differs from
prior work on cellular network architecture, scalable data center,
software defined networks, and middleboxes.

\myitem{Cellular network architecture:}
Recently work has exposed the complexity and inflexibility of current cellular
data networks~\cite{cell-directions, elby-sdn}. There are several
efforts~\cite{cell-directions, elby-sdn, CellSDNEuroSDN12, cloudEPC12,
  OpenRoads:Visa2010} attempting to fix the problem. However,
only~\cite{OpenRoads:Visa2010, cloudEPC12} have concrete designs.
OpenFlow Wireless~\cite{OpenRoads:Visa2010}
focuses on virtualizing data path and configuration. ~\cite{cloudEPC12} proposes
an integration of OpenFlow with LTE control plane so that GTP tunnels can be
setup using OpenFlow. 
None of them present scalable network architecture for fine-grained policy.

\myitem{Scalable data centers:} Our addressing scheme shares some similarity to
prior work on scalable data center. VL2~\cite{VL2-09} assigns servers IP
addresses that act as names alone. 
PortLand~\cite{Portland09} assigns internal Pseudo MAC
addresses to all end hosts to encode their position in the
topology. Nicira~\cite{Nicira}'s virtual data center networks require
intelligent Internet gateways. Our gateway switches are much simpler because we
``embed'' policy and location information in the packet header, rather
than relying on the controller to install fine-grain packet-classification
rules. 

\myitem{Software defined networks:} Recent work~\cite{devoflow,difane} improves
upon Ethane~\cite{ethane} to avoid maintaining per micro flow state in switches.
DevoFlow~\cite{devoflow} which handles most micro-flow rules in the data
plane. DIFANE~\cite{difane} distributes pre-installed OpenFlow wildcard rules
among multiple switches and ensures all decisions can be made in the data-plane. 
Unlike SoftCell, they do not support policy policy symmetry and policy
consistency. 
SoftCell architecture conforms to~\cite{SDNInternet12}. SoftCell
distinguishes edge from core. The core routes on tags and IP prefixes that are
different from the UE addresses. In addition, SoftCell differentiates access
edge from gateway edge. SoftCell minimizes state kept at gateway switches. 

\myitem{Middleboxes:} Prior work has focused on (1) middlebox design, e.g. a single
box with modular capabilities that can implement a vast variety of services (for
instance, see~\cite{sdnMB12,manefesto11}); (2) mechanisms to
enforce middlebox traversals~\cite{pswitch08}. However,
they do not present any scalable network architecture for fine-grained policy.

\section{Conclusion}
\label{sec:concl}
Today's cellular core networks are expensive and inflexible.  In this
paper, we propose SoftCell, a scalable architecture for supporting
fine-grained policies in cellular core networks.  SoftCell achieves
scalability in the data plane by (i) pushing packet classification to
low-bandwidth access switches and (ii) minimizing the state in core
network through effective, multi-dimensional aggregation of forwarding
rules. SoftCell achieves scalability in the control plane by caching
packet classifiers and policy tags at local agents that update the
rules in the access switches.  We further design a modular controller
that decouples high-level traffic management from the low-level
details of computing and installing switch-level rules.  Our prototype
and evaluation demonstrate that SoftCell significantly improves the
flexibility of future cellular core networks, while reducing cost
through the use of commodity switches and middleboxes.

\small
\bibliographystyle{ieeetr}
\bibliography{cellsdn}

\begin{thebibliography}{10}

\bibitem{cell-directions}
B.-j. Kim and P.~Henry, ``Directions for future cellular mobile network
  architecture,'' {\em First Monday}, vol.~17, no.~12, 2012.

\bibitem{elby-sdn}
S.~Elby, ``Carrier vision of {SDN} and future applications to achieve a more
  agile mobile business,'' October 2012.
\newblock Keynote address at the SDN \& OpenFlow World Congress,
  \url{http://www.layer123.com/sdn-live}.

\bibitem{nfv}
``Network functions virtualization: Introductory white paper,'' October 2012.
\newblock \url{http://www.tid.es/es/Documents/NFV_White_PaperV2.pdf}.

\bibitem{ethane}
M.~Casado, M.~J. Freedman, J.~Pettit, J.~Luo, N.~Gude, N.~McKeown, and
  S.~Shenker, ``Rethinking enterprise network control,'' {\em IEEE/ACM Trans.
  Networking}, vol.~17, August 2009.

\bibitem{load-wild}
R.~Wang, D.~Butnariu, and J.~Rexford, ``{OpenFlow}-based server load balancing
  gone wild,'' in {\em Hot-ICE Workshop}, March 2011.

\bibitem{past}
B.~Stephens, A.~Cox, W.~Felter, C.~Dixon, and J.~Carter, ``{PAST}: Scalable
  {Ethernet} for data centers,'' in {\em ACM SIGCOMM CoNext Conference},
  December 2012.

\bibitem{openvswitch}
``{Open vSwitch}.'' \url{http://openvswitch.org/}, 2013.

\bibitem{devoflow}
A.~R. Curtis, J.~C. Mogul, J.~Tourrilhes, P.~Yalagandula, P.~Sharma, and
  S.~Banerjee, ``{DevoFlow}: Scaling flow management for high-performance
  networks,'' in {\em ACM SIGCOMM}, August 2011.

\bibitem{celltopology}
M.~Howard, ``Using carrier ethernet to backhaul {LTE}.''
  \url{http://tinyurl.com/bdxl6wo}, 2011.
\newblock {Infonectics Research, White Paper}.

\bibitem{eNodeBALU}
Alcatel-Lucent, ``{Alcatel-Lucent} 9926 digital {2U eNode B}.''

\bibitem{rahmati}
A.~Rahmati, C.~Shepard, C.~Tossell, A.~Nicoara, L.~Zhong, P.~Kortum, and
  J.~Singh, ``Seamless flow migration on smartphones without network support,''
  {\em IEEE Transactions on Mobile Computing}, 2013.
\newblock To appear.

\bibitem{zhang12}
Y.~Zhang and A.~Arvidsson, ``Understanding the characteristics of cellular data
  traffic,'' in {\em ACM SIGCOMM CellNet Workshop}, August 2012.

\bibitem{Ericsson1G}
L.~Whitney, ``Ericsson demos faster lte speeds of almost {1Gbps}.''
  \url{http://tinyurl.com/alml5vt}.

\bibitem{ovs-perf}
``The rise of soft switching, part {II}: Soft switching is awesome,'' June
  2012.
\newblock \url{http://tinyurl.com/bjz8469}.

\bibitem{oflops}
C.~Rotsos, N.~Sarrar, S.~Uhlig, R.~Sherwood, and A.~W. Moore, ``{OFLOPS}: An
  open framework for {OpenFlow} switch evaluation,'' in {\em Workshop on
  Passive and Active Measurement}, March 2012.

\bibitem{wang2011untold}
Z.~Wang, Z.~Qian, Q.~Xu, Z.~Mao, and M.~Zhang, ``An untold story of middleboxes
  in cellular networks,'' in {\em ACM SIGCOMM}, August 2011.

\bibitem{bscluster}
R.~Nadiv and T.~Naveh, ``Wireless backhaul topologies: Analyzing backhaul
  topology strategies,'' {\em Ceragon White Paper}, 2010.

\bibitem{controller-perf}
``Controller performance comparisons,'' May 2011.
\newblock
  \url{http://www.openflow.org/wk/index.php/Controller_Performance_Comparisons}.

\bibitem{HAWAII02}
R.~Ramjee, K.~Varadhan, L.~Salgarelli, S.~Thuel, S.-Y. Wang, and T.~La~Porta,
  ``Hawaii: A domain-based approach for supporting mobility in wide-area
  wireless networks,'' {\em IEEE/ACM Trans. Networking}, vol.~10, June 2002.

\bibitem{CellularIP99}
A.~Campbell, J.~Gomez, and A.~Valko, ``An overview of cellular {IP},'' in {\em
  IEEE Wireless Communications and Networking Conference}, 1999.

\bibitem{MobileData18}
{Cisco}, ``Cisco visual networking index forecast projects 18-fold growth in
  global mobile internet data traffic from 2011 to 2016.''
  \url{http://tinyurl.com/7gn9x9s}.

\bibitem{floodlight}
``{Floodlight OpenFlow Controller}.''
\newblock \url{http://floodlight.openflowhub.org/}.

\bibitem{cbench}
``{Cbench OpenFlow Controller Benchmark}.''
\newblock \url{http://www.openflow.org/wk/index.php/Oflops}.

\bibitem{attIP}
AT\&T, ``Wireless {IP} options for mobile deployments.''
  \url{https://www.wireless.att.com/businesscenter/solutions/connectivity/ip-addressing.jsp}.

\bibitem{CellSDNEuroSDN12}
L.~Li, Z.~Mao, and J.~Rexford, ``Toward software-defined cellular networks,''
  in {\em European Workshop on Software Defined Networking (EWSDN)}, October
  2012.

\bibitem{cloudEPC12}
J.~Kempf, B.~Johansson, S.~Pettersson, H.~Luning, and T.~Nilsson, ``Moving the
  mobile evolved packet core to the cloud,'' in {\em IEEE WiMob}, October 2012.

\bibitem{OpenRoads:Visa2010}
K.-K. Yap, R.~Sherwood, M.~Kobayashi, T.-Y. Huang, M.~Chan, N.~Handigol,
  N.~McKeown, and G.~Parulkar, ``Blueprint for introducing innovation into
  wireless mobile networks,'' in {\em ACM VISA Workshop}, August 2010.

\bibitem{VL2-09}
A.~Greenberg, J.~R. Hamilton, N.~Jain, S.~Kandula, C.~Kim, P.~Lahiri, D.~A.
  Maltz, P.~Patel, and S.~Sengupta, ``{VL2}: A scalable and flexible data
  center network,'' in {\em ACM SIGCOMM}, August 2009.

\bibitem{Portland09}
R.~Niranjan~Mysore, A.~Pamboris, N.~Farrington, N.~Huang, P.~Miri,
  S.~Radhakrishnan, V.~Subramanya, and A.~Vahdat, ``{PortLand}: a scalable
  fault-tolerant layer 2 data center network fabric,'' in {\em ACM SIGCOMM},
  August 2009.

\bibitem{Nicira}
Nicira, ``It's time to virtualize the network: Network virtualization for cloud
  data centers.'' \url{http://tinyurl.com/c9jbkuu}.

\bibitem{difane}
M.~Yu, J.~Rexford, M.~J. Freedman, and J.~Wang, ``Scalable flow-based
  networking with {DIFANE},'' in {\em ACM SIGCOMM}, August 2010.

\bibitem{SDNInternet12}
B.~Raghavan, M.~Casado, T.~Koponen, S.~Ratnasamy, A.~Ghodsi, and S.~Shenker,
  ``Software-defined {Internet} architecture: Decoupling architecture from
  infrastructure,'' in {\em ACM SIGCOMM HotNets Workshop}, October 2012.

\bibitem{sdnMB12}
A.~Gember, P.~Prabhu, Z.~Ghadiyali, and A.~Akella, ``Toward software-defined
  middlebox networking,'' in {\em ACM SIGCOMM HotNets Workshop}, 2012.

\bibitem{manefesto11}
V.~Sekar, S.~Ratnasamy, M.~K. Reiter, N.~Egi, and G.~Shi, ``The middlebox
  manifesto: Enabling innovation in middlebox deployment,'' in {\em ACM SIGCOMM
  HotNets Workshop}, 2011.

\bibitem{pswitch08}
D.~Joseph, A.~Tavakoli, and I.~Stoica, ``A policy-aware switching layer for
  data centers,'' in {\em ACM SIGCOMM}, August 2008.

\end{thebibliography}

\end{document}